\documentclass[sigconf]{acmart}
\usepackage{subfig} 
\usepackage{xcolor}
\usepackage{fontawesome} 
\usepackage{pifont}
\usepackage{natbib}
\usepackage{enumitem} 
\usepackage{url} 
\usepackage{hyperref}
\usepackage{multirow}

\AtBeginDocument{%
  }

\setcopyright{acmcopyright}
\copyrightyear{2026}
\acmYear{2026}
\setcopyright{cc}
\setcctype{by}
\acmConference[MM '26]{Proceedings of the 35th ACM International Conference on Multimedia}{November 10--14, 2026}{Rio de Janeiro, Brazil}
\acmBooktitle{Proceedings of the 35th ACM International Conference on Multimedia (MM '26), November 10--14, 2026, Rio de Janeiro, Brazil}
\acmDOI{10.1145/3767308.3836628}
\acmISBN{979-8-4007-2213-4/2026/11}


\begin{document}

\title{AT-ADD: All-Type Audio Deepfake Detection Challenge Summary}

\author{Yuankun Xie}
\authornote{These authors contributed equally to this work. Official Website: \url{https://at-add.com}.}
\affiliation{%
  \institution{Communication University of China \& Ant Group}
  \city{Beijing}
  \country{China}
}

\author{Haonan Cheng}
\authornotemark[1]
\affiliation{%
  \institution{Communication University of China}
  \city{Beijing}
  \country{China}
}

\author{Jiayi Zhou}
\authornotemark[1]
\affiliation{%
  \institution{Machine Intelligence, Ant Group}
  \city{Shanghai}
  \country{China}
}

\author{Xiaoxuan Guo}
\affiliation{%
  \institution{Communication University of China}
  \institution{Ant Group}
  \city{Beijing}
  \country{China}
}

\author{Tao Wang}
\affiliation{%
  \institution{Machine Intelligence, Ant Group}
  \city{Shanghai}
  \country{China}
}

\author{Changhao Zhang}
\affiliation{%
	\institution{Machine Intelligence, Ant Group}
	\city{Shanghai}
	\country{China}
}

\author{Jian Liu}
\affiliation{%
  \institution{Machine Intelligence, Ant Group}
  \city{Shanghai}
  \country{China}
}

\author{Weiqiang Wang}
\affiliation{%
  \institution{Machine Intelligence, Ant Group}
  \city{Shanghai}
  \country{China}
}

\author{Ruibo Fu}
\affiliation{%
	\institution{Institute of Automation, Chinese Academy of Sciences}	
	\city{Beijing}
	\country{China}
}

\author{Xiaopeng Wang}
\affiliation{%
	\institution{Beijing Institute of Technology}
	\city{Beijing}
	\country{China}
}

\author{Hengyan Huang}
\affiliation{%
	\institution{Communication University of China}
	\city{Beijing}
	\country{China}
}

\author{Xiaoying Huang}
\affiliation{%
	\institution{Communication University of China}
	\city{Beijing}
	\country{China}
}
\author{Long Ye}
\affiliation{%
	\institution{Communication University of China}
	\city{Beijing}
	\country{China}
}

\author{Guangtao Zhai}
\affiliation{%
	\institution{Shanghai Jiao Tong University}
	\city{Shanghai}
	\country{China}
}

\renewcommand{\shortauthors}{Yuankun Xie et al.}

\begin{abstract}
This paper summarizes the ACM Multimedia 2026 AT-ADD Grand Challenge on all-type audio deepfake detection. AT-ADD contains two tracks: robust speech deepfake detection under realistic acoustic and channel variations, and type-agnostic detection over speech, environmental sound, singing voice, and music. We describe the challenge tasks, dataset and evaluation-set design, official leaderboard results, and common design patterns observed in participating systems. The best Track~1 system achieved 90.71\% Macro-F1 on the final evaluation set, while the best Track~2 system achieved 96.10\% Macro-F1. The final submissions show that strong systems commonly combine large-scale self-supervised audio representations, data augmentation, multi-crop inference, and structured fusion or routing. The results also reveal remaining challenges in generalization to unseen generators, robustness to realistic speech-domain distortions, and balanced performance across heterogeneous audio types.
\end{abstract}



\vspace{-3pt}


\maketitle
\vspace{-3pt}

\section{Introduction}

\begin{figure}[!tb]
	\centering
	\subfloat{\includegraphics[width=3.0in]{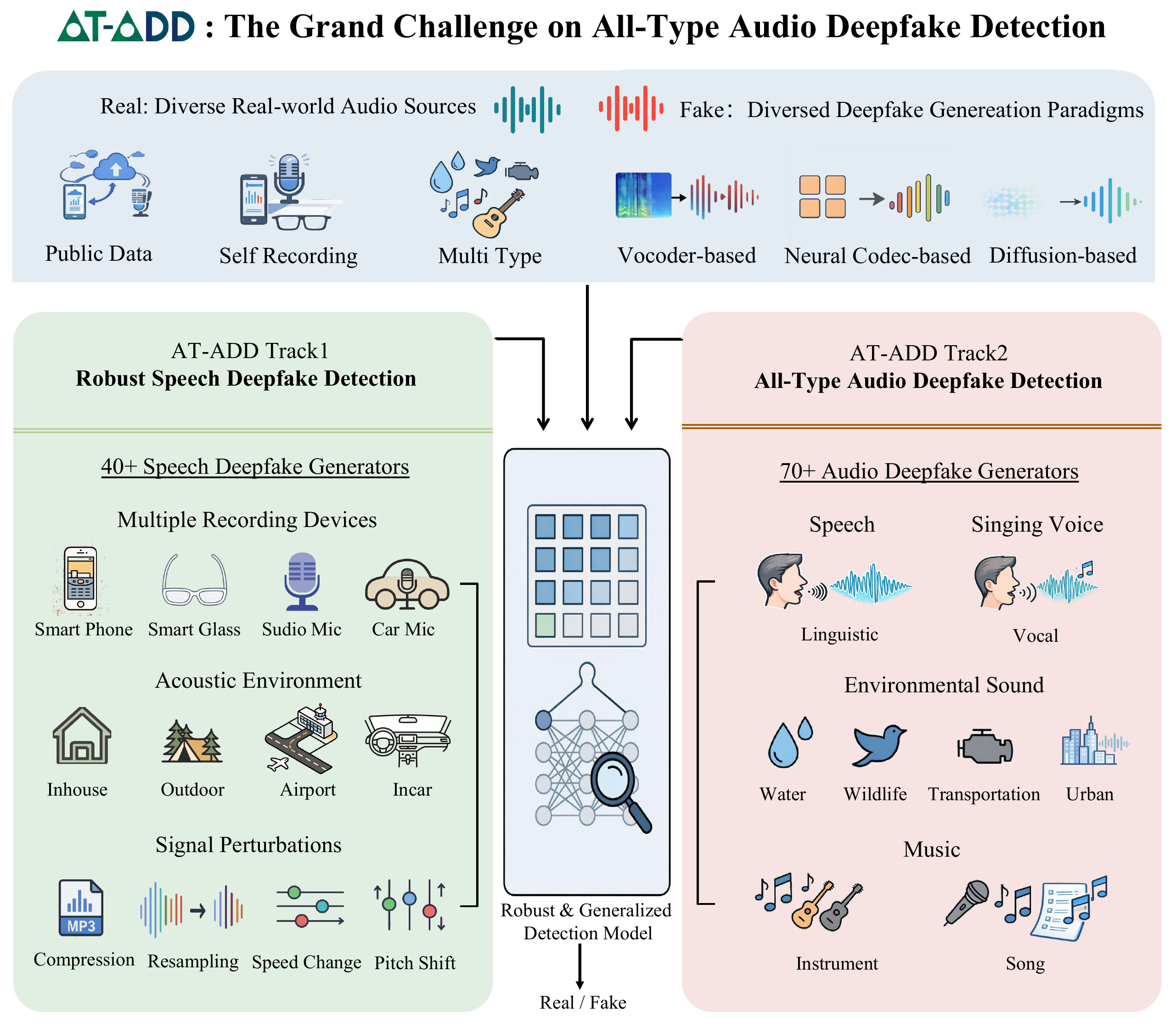}}
	\hfil
	\caption{AT-ADD challenge overview.}
	\Description{Overview diagram of the two AT-ADD tracks and their target audio types.}
	\label{fig:intro} 
\end{figure}

Recent audio generation models can synthesize high-fidelity speech, environmental sound, singing voice, and music. These capabilities create new risks for multimedia trust, since realistic fake audio can now appear in many content forms rather than only in text-to-speech or voice conversion. Existing audio deepfake detection (ADD) studies have advanced rapidly, especially through ASVspoof and ADD challenge series~\cite{wang2024asvspoof,yi2023add}, but practical deployment still faces two major gaps: robustness under real acoustic and channel conditions, and generalization across heterogeneous audio types.

The AT-ADD Grand Challenge addresses these gaps with two complementary tracks. Track~1, \emph{Robust Speech Deepfake Detection}, evaluates binary real/fake speech detection under unseen generators and realistic perturbations. Track~2, \emph{All-Type Audio Deepfake Detection}, extends the task to speech, sound, singing, and music, and participants must make a binary decision when the input audio type is unknown. Both tracks follow a closed setting: systems may use the official train/dev data, permitted signal-level augmentation, and traceable general-purpose pretrained models, but not external deepfake detection datasets or progress/evaluation-set adaptation.

\section{Tasks and Datasets}
\label{sec:tasks}

AT-ADD contains two complementary tracks under a \textbf{closed setting}, where participants train CMs only with organizer-provided data. This design enables controlled comparison of generalization and robustness under limited-data conditions.

\subsection{Track 1: Robust Speech Deepfake Detection}
\label{sec:track1}

\textbf{Goal.}
Track~1 aims to bridge the gap between existing benchmarks and real-world deployment scenarios for speech deepfake detection. It evaluates whether a detector can remain reliable under realistic domain shifts and practical post-processing effects, while maintaining strong performance against modern high-fidelity synthesis systems.

\textbf{Task definition.}
Given an input speech utterance, participants are required to predict whether the input is \emph{real} or \emph{fake}. In this task, \emph{fake} refers specifically to deepfake speech generated using deep neural network-based methods, while \emph{real} refers to non-deepfake speech. It should be noted that signal distortions or transformations, such as compression, resampling, speed perturbation, and pitch shifting, as well as replay-based attacks, do not change the original real/fake label in this task. The training and development data are fully provided by the organizers, and the use of external data is not allowed under the closed setting.

The evaluation set includes deepfake samples generated by methods that are \emph{unseen} during training and reflect recent state-of-the-art generation techniques. Meanwhile, the real speech in the evaluation set is collected under realistic conditions, involving variations in recording devices, acoustic environments, languages, and other real-world factors.

\subsection{Track 2: All-Type Audio Deepfake Detection}
\label{sec:track2}

\textbf{Goal.}
Track~2 targets universal audio deepfake detection across heterogeneous audio types and aims to develop \emph{type-agnostic} detectors that generalize across both audio types and unseen generation methods.

\textbf{Task definition.}
Given an input audio clip of unknown type, participants are required to determine whether it is \emph{real} or \emph{fake}. In this task, \emph{fake} denotes deepfake audio generated by deep neural network-based methods, whereas \emph{real} denotes non-deepfake audio. Notably, in Track~2, audio-type labels (i.e., speech, sound, singing, and music) are \emph{not} available at test time, reflecting realistic deployment scenarios.

Similar to Track~1, this track follows a closed setting, where participants must use only the provided training and development data, without access to external resources.

\subsection{Datasets and Resources}

We construct two benchmark datasets, \textbf{AT-ADD Track~1} and \textbf{AT-ADD Track~2}, with standardized train, development (dev), progress, and evaluation (eval) splits. The progress subset is sampled from the eval distribution and accounts for 20\% of the full eval set. Table~\ref{tab:alldata_protocol} summarizes the track statistics; minor count differences come from quality screening such as silent-segment removal.

\begin{table*}[t]
\centering
\caption{AT-ADD statistics (number of clips) for Track 1 and Track 2.}
\label{tab:alldata_protocol}
\begin{tabular}{c c | ccccc}
\hline
\multirow{2}{*}{\textbf{Split}} & 
\multirow{2}{*}{\textbf{T1}} & 
\multicolumn{5}{c}{\textbf{T2}} \\
\cline{3-7}
 &  & \textbf{Speech} & \textbf{Sound} & \textbf{Singing} & \textbf{Music} & \textbf{Total} \\
\hline
Train    & 49,575  & 49,575  & 39,840 & 36,000 & 21,366 & 146,781 \\
Dev      & 49,734  & 49,734  & 19,929 & 16,000 & 5,406  & 91,069 \\
Progress & 29,269  & 28,813  & 5,729  & 4,872  & 6,461  & 45,875 \\
Eval     & 146,346 & 144,078 & 28,593 & 24,332 & 32,370 & 229,373 \\
\hline
\end{tabular}
\end{table*}

\subsection{Track 1 Dataset Details}

We construct the AT-ADD Track~1 dataset with predefined training, development, and evaluation splits. The evaluation split is reserved for testing and consists of real speech collected from diverse domains, together with fake speech generated by methods that are unseen in the training and development sets, thereby enabling the evaluation of CM robustness under domain shifts, such as variations in recording devices, acoustic environments, and signal perturbations. 

\textbf{Real speech in train/dev sets.}
The real speech subset within the training and development sets is comprised of a diverse collection of multilingual utterances. This subset incorporates internal recordings captured across a variety of Recording devices to ensure acoustic diversity, alongside high-quality Chinese speech samples sourced from the AISHELL-3~\cite{shi2020aishell} dataset. To bolster the English portion, samples are integrated from the LibriTTS-R~\cite{koizumi2023libritts} and LJSpeech~\cite{ljspeech} corpora. Furthermore, the dataset's multilingual breadth is further extended through the inclusion of representative samples from Common Voice~\cite{ardila2020common}, covering a wide array of linguistic contexts.

\textbf{Fake speech in the train/dev sets.}
The fake speech subset covers multiple tasks, including text-to-speech (TTS) and voice conversion (VC). For TTS, the input texts are selected from the real speech data described in the previous subsection and used for synthesis. For VC and one-shot TTS tasks that require reference speaker cloning, we use the same pool of real reference speakers for the training and development sets, while a different pool of real reference speakers is used for the evaluation set to prevent speaker-information leakage. In addition, during cloning, we ensure that the reference speaker and the source speaker are never the same person.

\textbf{Real speech in Eval sets.}
The real speech were drawn from six source domains: 5k internally collected recordings captured using diverse recording devices, including smart glasses, mobile phones, and in-vehicle systems; 2k 3D-speaker~\cite{zheng20233d} recordings collected with different recording devices; 2k real non-replay samples from EchoFake~\cite{zhang2025echofake}; 2k samples from AISHELL-3; 4k samples from LibriTTS-R; and 5k samples from Common Voice. To evaluate robustness, 20\% of the real samples were subjected to signal perturbation, including 1k speed perturbations, 1k volume variations, 1k resampling operations, as well as 1k samples with various combinations of these transformations.

\textbf{Fake speech in Eval sets.}
The fake speech in the evaluation set is generated by 26 methods that are \emph{unseen} during training. In terms of synthesis paradigms, it covers several mainstream categories. A portion of the fake speech is further subjected to secondary transformations, including 5k samples with volume variations, 5k with resampling, 10k with speed perturbations, and 10k with combinations of these three transformations.

\subsection{Track 2 Dataset Details}

In this section, we describe the composition of the proposed AT-ADD Track~2 dataset across four different audio types. 

\textbf{Speech.}
We use the same speech training and development sets as in AT-ADD Track~1. However, the evaluation set is simplified by removing the signal perturbation and replay attak introduced in Track~1. As Track~2 targets universal, all-type deepfake detection, this design provides a clean and consistent evaluation protocol that emphasizes cross-type generalization rather than robustness to signal degradations.

\textbf{Sound.}
The sound subset is constructed from the AudioCaps dataset~\cite{kim2019audiocaps}. We first divide the audio samples in AudioCaps, which labeled as real samples, into non-overlapping training, development, and evaluation sets. The synthetic samples in the training and development sets are generated by TTA models conditioned on the corresponding textual descriptions of the real audio. For the evaluation set, the fake samples are generated from the remaining textual descriptions using 4 \emph{unseen} generation methods. In addition, we incorporate out-of-distribution (OOD) real audio samples from several public datasets—1K AVQA~\cite{yang2022avqa}, 1K CompA-R~\cite{ghosh2024gama}, 1K VocalSound~\cite{gong2022vocalsound}, and 1K TUT2016~\cite{mesaros2016tut}—into the evaluation set to assess the generalization capability of CMs.

\textbf{Singing Voice.}
The singing voice subset is constructed from three source datasets: OpenCpop~\cite{wang2022opencpop}, M4Singer~\cite{zhang2022m4singer}, and KiSing~\cite{shi2024singing}. As in the sound subset, the samples from these three source domains are first divided into non-overlapping training, development, and evaluation sets, which are labeled as the real samples. The fake samples in the training and development sets are generated via singing voice conversion, with strictly non-overlapping source and target singers to avoid identity leakage. The fake samples in the evaluation set are produced by 5 \emph{unseen} deepfake methods, allowing a comprehensive evaluation of cross-model generalization. 

\textbf{Music.}
The music subset is derived from the MusicCaps \cite{agostinelli2023musiclm} dataset. We first divide the audio samples in MusicCaps, which labeled as real samples, into non-overlapping training, development, and evaluation sets. The synthetic samples in the training and development sets are generated by TTM models conditioned on the corresponding textual descriptions of the real music. For the evaluation set, the fake samples are generated from the remaining textual descriptions using 4 \emph{unseen} generation methods. In addition, we incorporate OOD real music samples from public datasets—5K from the FMA dataset~\cite{defferrard2016fma} and 5K from FortisAVQA~\cite{ma2025fortisavqa}—into the evaluation set to assess the generalization capability of CMs.
\section{Final Results and Teams}

After the final leaderboard freeze, we verified submitted metadata and summarized the top five valid systems in Table~\ref{tab:final_results}. The table lists each team's main data augmentation strategy, system name, number of submitted subsystems, progress-set score, and final evaluation score.

\begin{table*}[t]
\centering
\caption{Final leaderboards for Track~1 robust speech deepfake detection and Track~2 all-type audio deepfake detection. Scores are Macro-F1 (\%).}
\label{tab:final_results}
\label{tab:t1_results}
\label{tab:t2_results}
\scriptsize
\setlength{\tabcolsep}{2.5pt}
\renewcommand{\arraystretch}{0.86}
\resizebox{\textwidth}{!}{
\begin{tabular}{llrlllrr}
\toprule
\textbf{Track} & \textbf{Rank} & \textbf{Team} & \textbf{Data Augmentation} & \textbf{System Name} & \textbf{Sub.} & \textbf{Progress} & \textbf{Eval} \\
\midrule
\multirow{5}{*}{T1} & 1 & WaveShield & Noise; Reverb; Codec; Perturb.; Replay; Seg. & w2vBERT2.0-AASIST & 3 & 84.54 & \textbf{90.71} \\
 & 2 & Fosafer & Noise; Reverb; Perturb.; Seg. & ssl\_gfcc\_multiscale\_ensemble & 4 & 85.31 & 86.67 \\
 & 3 & sonomsl & Noise; Reverb; Codec; Perturb.; Replay & sonomsl & 3 & \textbf{86.25} & 86.63 \\
 & 4 & ThreeTO & Codec; RB; Perturb.; Replay & atadd\_track1\_xlsr\_aasist\_ensemble & 3 & 83.61 & 83.79 \\
 & 5 & NKU-HLT & Reverb; Codec & EnvRobust-XLSR-AASIST & 1 & 84.20 & 83.68 \\
\midrule
\multirow{5}{*}{T2} & 1 & starfire & Codec; RB; Seg. & starfire\_track2\_audio\_type\_routed\_ensemble & 4 & \textbf{96.28} & \textbf{96.10} \\
 & 2 & orange9 & Noise; Reverb; Codec; RB; Perturb. & E2E-ATADD & 4 & 95.13 & 95.58 \\
 & 3 & ThreeTO & RB & MEDS & 3 & 93.76 & 93.95 \\
 & 4 & Fosafer & Noise; Reverb; Perturb.; Seg. & ssl\_gfcc\_scene\_adaptive\_ensemble & 5 & 91.61 & 91.62 \\
 & 5 & KETI-KU & Codec & Wav2Vec2-BERT-AASIST & 1 & 91.31 & 91.10 \\
\bottomrule
\end{tabular}}
\begin{minipage}{\textwidth}
\vspace{1mm}
\footnotesize\emph{Note:} Noise denotes augmentation using MUSAN or other public noise corpora. Reverb denotes reverberation-related augmentation, e.g., RIR. Codec denotes codec or re-encoding augmentation, e.g., g711alaw, mp3, opus, etc. Perturb. denotes signal-level perturbations, e.g., quantization, dynamic-range compression, clipping, EQ, masking, resampling, time stretching, pitch shifting, etc. Replay denotes playback simulation through loudspeaker playback. Seg. denotes segment-level operations, e.g., random crop and bonafide segment concatenation. RB denotes RawBoost.
\end{minipage}
\end{table*}

\subsection{Track 1 Team Systems}

The top Track~1 systems all relied on strong speech self-supervised learning (SSL) representations, but differed in how they controlled channel and generator mismatch. WaveShield ranked first with a three-member W2V-BERT~2.0\footnote{\url{https://huggingface.co/facebook/w2v-bert-2.0}} ensemble using AASIST~\cite{jung2022aasist}, AASIST3~\cite{borodin2024aasist3}, and Adapter-MFA back-ends. Each subsystem was trained with a staged frozen-LoRA~\cite{hu2022lora} to joint fine-tuning schedule, AM-Softmax plus focal loss, and broad condition augmentation, including MUSAN noise~\cite{David2015MUSAN}, RIR reverberation~\cite{Tom2017A}, codec or re-encoding artifacts, signal perturbation, replay simulation, and segment-level bonafide construction. The submitted score was produced by averaging subsystem logits followed by a fixed threshold.

Fosafer ranked second with a four-subsystem multi-scale XLSR~\cite{babu2021xls} ensemble. The system used XLSR front-ends at 0.3B\footnote{\url{https://huggingface.co/facebook/wav2vec2-xls-r-300m}}, 1B, and 2B scales, TSSDNet/ATSSDNet back-ends~\cite{hua2021tssdnet}, and an additional GFCC-enhanced branch. Its online augmentation covered noise, background music, environmental sounds, room reverberation, level variation, filtering, silence insertion, and speed/pitch perturbation; 30\% of the official training samples were randomly augmented online, and final decisions were made by score-level fusion with a fixed threshold.

sonomsl ranked third with three variants of a common WPT-XLSR1B-AASIST architecture. The system used a mostly frozen XLSR-1B front-end with wavelet prompt tuning~\cite{xie2025detect} and an AASIST graph-attention back-end. Its three members introduced class rebalancing, condition-matched real/fake pairing, and partial fine-tuning of the last six XLSR layers. Per-window logits from 4.04-second sliding windows were aggregated by median pooling, fused with logistic regression, and regularized through DANN~\cite{ganin2015unsupervised} and GroupDRO~\cite{sagawa2020groupdro} to reduce generator and condition shortcuts.

ThreeTO ranked fourth with a three-subsystem wav2vec2-XLSR-AASIST ensemble. Its first branch fused XLSR layers 3, 11, and 24 and replaced the original AASIST input projection with a GRKAN module~\cite{yang2024kat}. The second branch added CQCC~\cite{todisco2016new} features through SSL-to-CQCC cross-attention, while the third used the same CQCC-enhanced design with stronger augmentation, including codec round trips, RawBoost~\cite{tak2022rawboost}, pitch shift, resampling, speed perturbation, and low-probability fake-only replay. The final score was a weighted logit fusion of the three branches.

NKU-HLT ranked fifth with a compact single EnvRobust-XLSR-AASIST system. It used wav2vec2-XLSR-300m with AASIST and a two-stage optimization strategy. The first stage fully fine-tuned the model with codec-resampling augmentation over internet and telephony codecs, while the second stage added RIR views with the SSL front-end frozen. A score-consistency mean squared error (MSE) term forced RIR-view scores to match clean-view scores, making the single model more robust to reverberant and codec-shifted speech conditions.

\subsection{Track 2 Team Systems}

Track~2 required a binary decision when the input type could be speech, sound, singing, or music. The winning starfire system used hard audio-type routing: a frozen BEATs~\cite{chen2023beats} router classified each clip into one of the four types, and a branch-specific detector then produced the real/fake decision. The speech branch used wav2vec2-XLSR with AASIST, while the non-speech branches relied on EAT-large~\cite{chen2024eat} AASIST detectors trained with music- and sound-biased data compositions. The final rules used class-specific thresholds and conservative multi-crop pooling rather than a single global score average.

orange9 ranked second with E2E-ATADD, a unified end-to-end detector that fused EAT-large and wav2vec2-XLSR-300m representations. Hidden layers from both front-ends were aggregated with learnable softmax weights, concatenated, passed through SwiGLU~\cite{shazeer2020glu}, and pooled by multi-head attention before multilayer perceptron (MLP) classification. The team trained on merged train/dev data using stratified five-fold splits, RawBoost augmentation, weighted cross-entropy, random 4-second crops, and five-crop inference. Auxiliary modules included Whisper-large-v3~\cite{radford2023whisper}  speech routing, an XLSR speech refiner trained with m4a re-encoding, MUSAN, and RIRS\_NOISES, and a parallel type-classification head.

ThreeTO ranked third with Multi-Expert Audio Deepfake detection with dual Self-supervised learning frontends (MEDS), a multi-expert detector built around XLSR and BEATs. The main system fused sparse SSL layers from XLSR (3, 11, 24) and BEATs (3, 6, 9), then used a five-branch AASIST classifier with four type-specialist branches and one unified branch. A second MEDS-CQCC subsystem added CQCC-like frequency-domain features through an eight-head SSL-to-CQCC cross-attention module. An auxiliary XLSR audio-type router selected type-specific aggregation and thresholds, and RawBoost was applied selectively to sound and singing samples.

Fosafer ranked fourth with a scene-adaptive ensemble. It combined four real/fake detectors based on XLSR front-ends of different scales, TSSDNet/ATT-TSSDNet back-ends~\cite{hua2021tssdnet}, and a GFCC-enhanced branch, together with an EAT-large four-class scene classifier. During training, 30\% of the official training samples were randomly augmented online with noise, reverberation, level variation, filtering, silence insertion, and speed/pitch perturbation. The final decision fused subsystem scores and then applied scene-specific thresholds predicted by the classifier.

KETI-KU ranked fifth with a single W2V-BERT-AASIST system. Unlike the larger routed or ensemble systems, it used one fully fine-tuned W2V-BERT~2.0 front-end with an AASIST back-end and weighted cross-entropy. A development-set threshold was then used for the final binary decision.

\section{Discussion and Conclusion}
The final results show that competitive systems rely on pretrained SSL front-ends, augmentation, multi-crop inference, and calibrated fusion, consistent with recent ADD findings~\cite{tak2022automatic,jung2022aasist,xie2025detect}. Track~1 remains particularly challenging because unseen generators, realistic acoustic and channel conditions, and limited labeled data introduce substantial distribution shifts, making augmentation and domain-robust learning important directions for further improvement. Meanwhile, Track~2 presents a different challenge: overall performance can hide uneven difficulty across audio types, requiring type-aware decisions and specialized modeling strategies to handle heterogeneous artifacts among speech, sound, singing, and music.
Future editions should report finer breakdowns by generator, perturbation, audio type and reproducibility details. AT-ADD thus provides both a leaderboard and a benchmark for practical audio deepfake detection.
\begin{acks}
This work was supported by the Beijing Natural Science Foundation (L252143), Fundamental Research Funds for the Central Universities (CUC26TD03), and Ant Group Research Intern Program.
\end{acks}
\bibliographystyle{ACM-Reference-Format}
\balance
\bibliography{myrefs}

\end{document}